\documentclass[11pt]{article}

\usepackage[a4paper,margin=1in]{geometry}
\usepackage{amsmath,amssymb,amsfonts}
\usepackage{graphicx}
\usepackage{bm}
\usepackage{hyperref}
\usepackage{microtype}
\usepackage{caption}
\usepackage{subcaption}
\usepackage{booktabs}
\usepackage{tcolorbox}
\tcbuselibrary{listings}  
\tcbuselibrary{breakable} 

\title{\bf Quantum Functional Information through the Evolution of Random Circuits}

\author{Rodrigo Pasti and Jonas Krause\\
rodrigo.pasti@pucpr.br and jonas.krause1@pucpr.br\\
Postgraduate Program in Smart and Sustainable Cities (PPGCIS)\\
Pontifical Catholic University of Paraná (PUCPR)}

\newcommand{\tr}{\operatorname{Tr}}
\newcommand{\ket}[1]{\left|#1\right\rangle}
\newcommand{\bra}[1]{\left\langle#1\right|}

\begin{document}
\maketitle

\begin{abstract}
We introduce Quantum Functional Information (QFI), a new metric to quantify the rarity and utility of quantum states and circuits. Unlike standard measures such as fidelity or entropy, QFI captures the balance between functionality and rarity within the Hilbert space. We validate QFI through two approaches: random circuit sampling  and evolutionary algorithms guided by fidelity or QFI objectives. Random sampling shows that states with near-perfect fidelity are less informational than slightly suboptimal but rarer states, while correlations reveal that high fidelity typically requires fewer gates and shallower circuits. Evolutionary results demonstrate that fidelity-only optimization converges quickly but reduces diversity and robustness, whereas QFI optimization sustains exploration, generates richer structures, and favors robust circuits with high fidelity. These findings position QFI as a practical and interpretable tool for circuit design, benchmarking, variational quantum algorithms, and exploring emergent patterns in quantum systems.
\end{abstract}

\section{Introduction}

Information is one of the most fundamental concepts across science, underpinning disciplines from physics and biology to computation and communication. In its classical sense, information is often defined in terms of the reduction of uncertainty, as formulated by Shannon. This view highlights information as a measurable quantity that captures the statistical properties of signals or states, regardless of their specific meaning. In quantum computing, however, information acquires a deeper dimension. Quantum information is encoded in the state of quantum systems, represented in Hilbert spaces and manipulated through the principles of superposition, entanglement, and measurement. This extension of the concept reveals that information is an intrinsic property of physical systems.

Within this broader framework, the notion of functional information has emerged as an attempt to move beyond the statistical description of data. Functional information  focuses on the capacity of information to perform a function or achieve a purpose. In natural sciences, it has been used to quantify the rarity of configurations that accomplish a particular function, such as in biological evolution or chemical self-organization \cite{Szostak2003FunctionalInfo,Hazen2007FunctionalInfo,Wong2023FunctionalSelection,Smith2020FunctionalConstraints,Pasti2018Biogeographic}. In other words, functional information highlights the relationship between structure and utility, distinguishing configurations that simply exist from those that are effective in a given context.

Building upon these foundations, we introduce the concept of Quantum Functional Information (QFI). From a mathematical perspective, QFI aims to provide a framework to quantify the rarity and functionality of quantum states or circuits within the exponentially large Hilbert space. While conventional quantum information theory often deals with entropy measures, correlations, or capacities, QFI asks: which quantum states are not only possible but also functional under specific tasks or constraints? From an application perspective, QFI can help identify and characterize states that are valuable for computation, communication, or simulation, such as maximally entangled states or rare configurations that provide computational advantage, like minimizing noise in NISQ computers \cite{Chen2023ComplexityNISQ}. By linking rarity, structure, and usefulness in the quantum domain, the study of Quantum Functional Information opens a new direction for both theoretical exploration and practical applications.

The primary objective of this article is to take the first steps toward the formulation of a functional information metric for the evaluation of quantum systems. While information-theoretic measures such as entropy, mutual information, and entanglement entropy \cite{Aifer2023ThermoQI,Hsieh2025DynamicalLandauer,Xu2025SeparableInfoMeasures} have been widely studied in quantum information science, they do not directly capture the functional role of quantum states or circuits. Our proposal is to construct a framework that not only quantifies how much information a system contains, but also assesses the degree to which that information is rare, structured, and useful for a given quantum task. 

To begin this exploration, we concentrate on the generation of  quantum circuits, where the structure and functionality of quantum states can be studied. Two complementary methodologies were adopted to evaluate the proposed metric. The first approach is based on the random generation of quantum circuits, with the aim of observing the spontaneous emergence of functional states. In this context, functionality is defined in terms of a target property, such as fidelity with a desired quantum state that allows us to investigate how frequently and under what conditions functional configurations appear within a vast space of possibilities.

The second approach employs an evolutionary algorithm designed to guide the exploration of the circuit space more efficiently. Rather than relying solely on random sampling, the evolutionary strategy adapts to progressively search for circuits that maximize functionality under the proposed metric. This method enables the discovery of a broader variety of circuits that fulfill specific criteria, thereby expanding the landscape of candidate functional configurations.

Together, these approaches provide an initial validation of the functional information metric in the quantum domain. Random circuit sampling reveals the baseline rarity of functional states \cite{HangleiterEisert2023,AharonovGaoLandauLiuVazirani2023}, while evolutionary search demonstrates how guided processes can enhance their discovery. By integrating these results, we establish the first step for the development of a general framework of QFI, opening pathways for future applications in quantum algorithm design, benchmarking, and the study of emergent patterns in complex quantum systems \cite{Estrada2024WhatIsComplexSystem}.

Our contributions are threefold. First, we provide a precise algebraic definition of QFI and prove that it is bounded, interpretable, and tunable. Second, we conduct experiments using Qiskit-based simulations where QFI guides evolutionary circuit searches. Third, we present findings, insights, and interpretations of the results, showing how QFI both captures intuitive properties of entangled circuits and reveals structural trade-offs in quantum systems.

The text is organized as follows:

\begin{itemize}
    \item \textbf{Introduction} – Presents the motivation for functional information in quantum systems and introduces the concept of Quantum Functional Information (QFI).  
    \item \textbf{Functional Information} – Reviews the classical definition of functional information, its origins in biology, and its mathematical formulation.  
    \item \textbf{Quantum Functional Information} – Extends the concept to quantum circuits and states, defining QFI in terms of rarity and functionality in Hilbert space.  
    \item \textbf{Empirical Determination of QFI} – Describes the random sampling of millions of circuits, the estimation of fidelity distributions, and the computation of QFI curves.  
    \item \textbf{Evolution of QFI} – Applies evolutionary algorithms to evolve circuits under different objectives (fidelity vs QFI), comparing their optimization behavior.  
    \item \textbf{Conclusion} – Summarizes the findings.  
\end{itemize}

\section{Functional Information}

The concept was introduced in molecular biology by Jack Szostak \cite{Szostak2003FunctionalInfo} and later formalized by Robert Hazen and colleagues in the context of the origins of life \cite{Hazen2007FunctionalInfo}. The motivation was to understand why some molecular arrangements (like proteins, RNA, or other polymers) are able to perform specific functions, such as catalyzing a reaction, while most random arrangements do not.
What does Functional Information (FI) measure?

\begin{itemize}
    \item The space of all possible configurations (for example, all possible protein sequences).
    \item The subset of configurations that achieve a function (for example, sequences that fold properly and catalyze a reaction).
    \item The rarity of those functional sequences within the vast space of all possible ones.
\end{itemize}

Therefore, FI quantifies how rare or specialized a configuration is within a given search space when it satisfies a defined function. Consider a system with a search space of possible configurations 
\(\mathcal{C}\), each configuration \(c \in \mathcal{C}\) being evaluated by a 
function \(f(c)\) that measures its performance or ability to achieve a predefined goal.

Let us define:

\begin{itemize}
    \item \(P(F)\): the probability that a randomly chosen configuration \(c\) achieves at least a given functional threshold \(T\). That is,
    \[
    P(F) = \Pr[f(c) \geq T].
    \]
    \item FI is then the negative logarithm (base 2, measuring in bits) of this probability:
    \[
    I(F) = - \log_{2} P(F).
    \]
\end{itemize}

then:

\[
I(F) = - \log_{2} \Pr[f(c) \geq T]
\]

where:
\begin{itemize}
    \item \(f(c)\) = functional performance of configuration \(c\),
    \item \(T\) = functional threshold,
    \item \(\Pr[f(c) \geq T]\) = probability that a random configuration satisfies the function above threshold \(T\).
\end{itemize}

It is possible to conclude:

\begin{itemize}
    \item If many configurations achieve the function (\(P(F)\) large), then \(I(F)\) is small --- little functional information.
    \item If only rare configurations achieve the function (\(P(F)\) small), then \(I(F)\) is large --- high functional information.
\end{itemize}

Now suppose each configuration \(c \in \mathcal{C}\) produces a measurable value 
\(f(c)\). Let \(v\) be a specific target value of interest.

\begin{itemize}
    \item \(P(f(c) = v)\): the probability that a randomly chosen configuration yields exactly the value \(v\).
\end{itemize}

Then, the FI associated with this value is:

\[
I(v) = - \log_{2} \Pr[f(c) = v].
\]

It is possible to conclude:

\begin{itemize}
    \item If the value \(v\) occurs frequently (\(P(f(c) = v)\) large), then \(I(v)\) is small.
    \item If the value \(v\) is rare (\(P(f(c) = v)\) small), then \(I(v)\) is large.
\end{itemize}

Thus, FI measures the informational significance of obtaining a specific value of the functional evaluation. In other words, FI is a measure of the rarity of functionality within a search space.

\section{Quantum Functional Information}

Functional information provides a quantitative measure of how rare or specialized certain quantum states are within the enormous Hilbert space of possibilities. 
Canonical examples include entangled states such as Bell states and Greenberger–Horne–Zeilinger (GHZ) states. 
Although many random quantum states exhibit some degree of entanglement, highly structured entangled states such as Bell and GHZ states occupy only a very small fraction of Hilbert space. 
Therefore, when such states emerge from a quantum circuit, they carry a large amount of functional information: they are rare outcomes that nevertheless encode important functionality for quantum tasks such as teleportation \cite{Acin2023BellFoundations}, superdense coding \cite{CharacterizationGHZ2023}.

Consider the search problem in Hilbert space: given an initial state, the aim is to find a \emph{target functional state}, for example, an entangled state of two or more qubits. 
The full Hilbert space of $n$ qubits has dimension $2^n$, and only a small subset of this space corresponds to highly functional states with useful entanglement properties. 
Thus, functional information can be used to measure the informational significance of finding such a state relative to  all possible states.

Let $C$ denote the set of possible configurations corresponding to quantum circuits. Each element $c \in C$ represents a circuit which, when applied to an initial state, produces an output quantum state.  More generally, one can view $C$ as the set of all possible transformations of quantum states achievable with a chosen universal gate set. We then define a functional evaluation:
\[
f(c): C \to \mathbb{R},
\]
where $f(c)$ produces a measurable quantity associated with the circuit $c$.  This can be, for example, the fidelity of the circuit’s output state with respect to a target entangled state, the amount of entanglement entropy generated, or another functional criterion of interest.

The task is then to obtain the probability distribution of $f(c)$ across the set $C$. 
Returning to the general definition of functional information, the quantum functional information associated with a condition $f(c) \geq T$ is:

\[
I(F) = - \log_{2} \Pr[f(c) \geq T],
\]

where $T$ is the functional threshold of interest. 
This formulation captures the rarity of quantum circuits that generate functional outcomes, thereby quantifying their informational value.

In the next section, we will empirically determine the distribution of $f(c)$ by sampling ensembles of quantum circuits. 
This will allow us to compute quantum functional information for specific cases, such as the emergence of Bell or GHZ states.

\section{Empirical Determination of Quantum Functional Information}

In this article, we adopt fidelity as the functional evaluation of a quantum circuit. 
For a given configuration \(c \in C\), the function \(f(c)\) is defined as the fidelity between the output state \(\rho_c\) of the circuit and a fixed reference state \(\sigma\):

\[
f(c) = F(\rho_c, \sigma),
\]

We compute the fidelity with a pure target $\tau$. For $n=2$ we use the Bell state $\ket{\Phi^+} = (\ket{00}+\ket{11})/\sqrt{2}$, while for $n \geq 3$ the target is the GHZ state $\ket{\mathrm{GHZ}_n} = (\ket{0}^{\otimes n}+\ket{1}^{\otimes n})/\sqrt{2}$. The fidelity is
\begin{equation}
F(\rho_C,\tau) = \bra{\tau}\rho_C\ket{\tau}.
\end{equation}

If both the output state of the circuit and the reference state are pure, then 
\(\rho_c = |\psi_c\rangle \langle \psi_c|\) and 
\(\sigma = |\phi\rangle \langle \phi|\). 
In this case, the fidelity reduces to the squared inner product between the states:

\[
F(\rho_c, \sigma) = \big| \langle \phi \,|\, \psi_c \rangle \big|^2.
\]

This choice allows us to quantify the probability of obtaining output states close to the reference state, which can be, for example, a Bell or GHZ state. If \(f(c)\) represents the fidelity of circuit \(c\), then the functional information can be expressed in terms of the probability distribution of fidelity across the circuit ensemble. Thus, the task reduces to obtaining the empirical distribution of fidelities over the configuration set \(C\).

To estimate this distribution, we performed a random sampling of quantum circuits with up to 50 gates. The gate set included both single-qubit and two-qubit gates:

\begin{itemize}
    \item Identity: $I$
    \item Pauli-X (NOT): $X$
    \item Pauli-Y: $Y$
    \item Pauli-Z: $Z$
    \item Hadamard: $H$
    \item Phase: $S$, $S^\dagger$
    \item $\pi/8$ gate: $T$, $T^\dagger$
    \item Rotation-X: $R_x(\theta)$
    \item Rotation-Y: $R_y(\theta)$
    \item Rotation-Z: $R_z(\theta)$
    \item Phase rotation: $R_{\phi}(\lambda)$
    \item Controlled-NOT (CNOT): $\mathrm{CX}$
    \item Controlled-Z: $\mathrm{CZ}$
    \item Controlled-Y: $\mathrm{CY}$
\end{itemize}

\begin{figure}[h]
    \centering
    \begin{subfigure}{0.48\textwidth}
        \includegraphics[width=\linewidth]{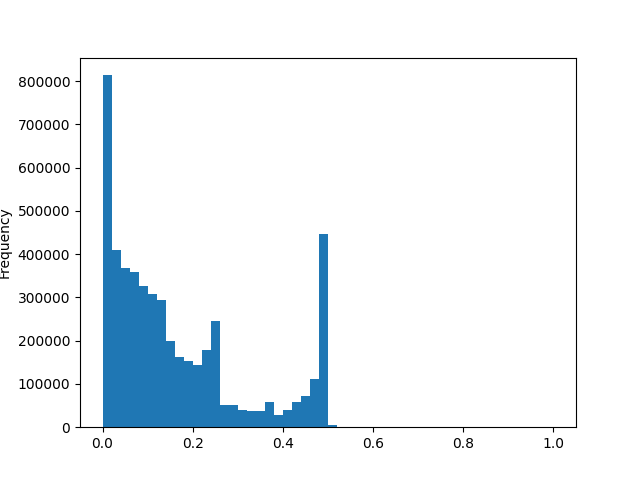}
        \caption{$n=4$ qubits}
    \end{subfigure}\hfill
    \begin{subfigure}{0.48\textwidth}
        \includegraphics[width=\linewidth]{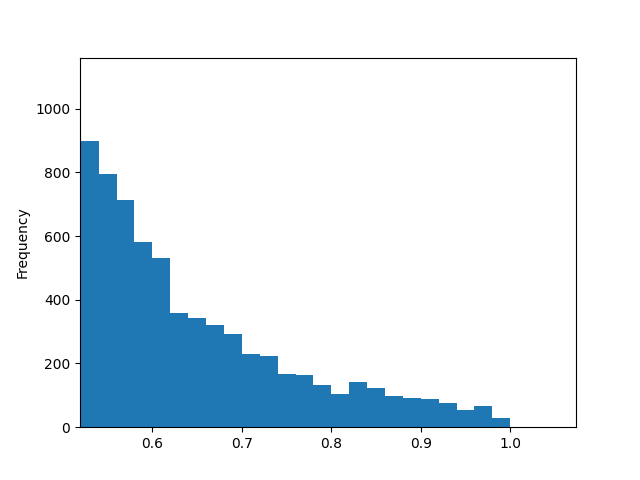}
        \caption{$n=4$ qubits}
    \end{subfigure}

    \begin{subfigure}{0.48\textwidth}
        \includegraphics[width=\linewidth]{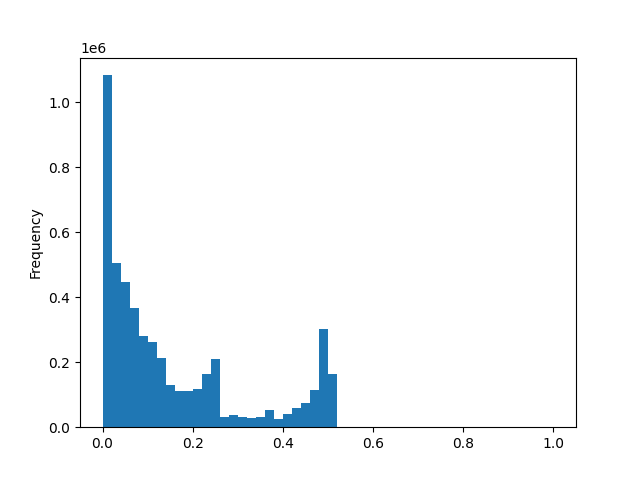}
        \caption{$n=5$ qubits}
    \end{subfigure}\hfill
    \begin{subfigure}{0.48\textwidth}
        \includegraphics[width=\linewidth]{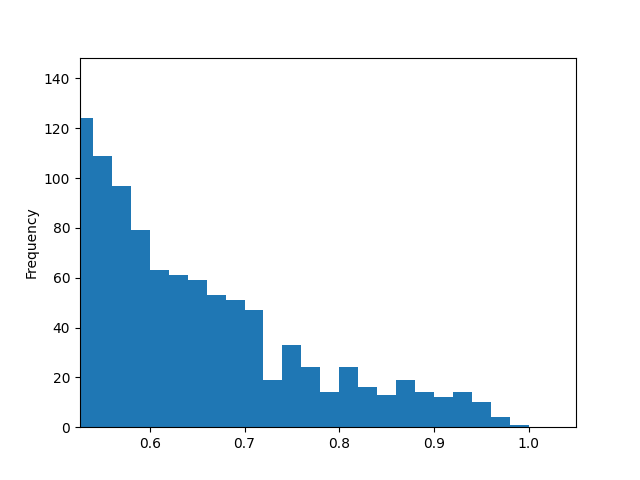}
        \caption{$n=5$ qubits}
    \end{subfigure}
 
    \caption{Random sampling of circuits and the relative frequency of fidelity. Full (a) and partial (b) distribution for 4 qubits and full (c) and partial (d) distribution for 5 qubits.}
    \label{fig:fid_histogram}
\end{figure}

A total of 5 million random circuits were generated for circuits with 2, 3, 4, and 5 qubits. For each circuit, the fidelity \(f(c)\) with respect to the target state was computed. 
The resulting histograms (Figure~\ref{fig:fid_histogram}) summarize the observed fidelity distribution. A large fraction of circuits display fidelity values below 0.5, while fewer circuits achieve higher fidelity, reflecting the increasing rarity of functional states.

To obtain a smooth estimate of the underlying probability distribution, the sampled data were discretized into 200 intervals. 
For each interval, the mean fidelity value and the corresponding probability were computed. 
These points served as the training data for nonlinear regression. 

Several regression algorithms were evaluated, each with different trade-offs in terms of bias, variance, and sensitivity to outliers. 
The results presented here were obtained using a decision tree regression, which achieved a high \(R^2\) score while avoiding interpolation of outlier points.

The fitted distribution was then used to compute quantum functional information by applying the equation

\[
I(f) = - \log_{2} P(f).
\]

\begin{figure}[h]
    \centering
    \begin{subfigure}{0.48\textwidth}
        \includegraphics[width=\linewidth]{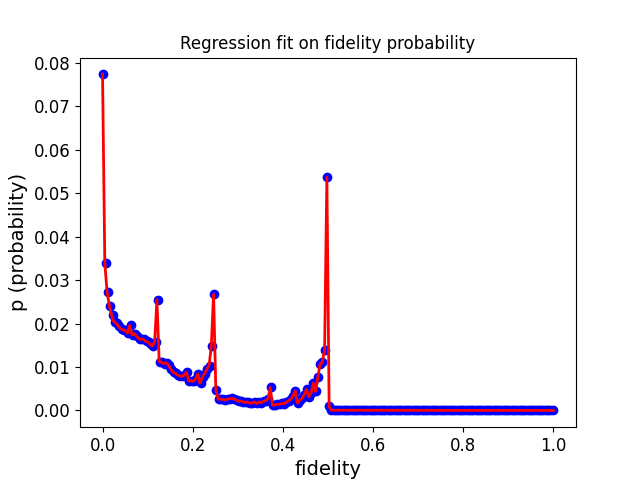}
        \caption{$n=4$ qubits}
    \end{subfigure}\hfill
    \begin{subfigure}{0.48\textwidth}
        \includegraphics[width=\linewidth]{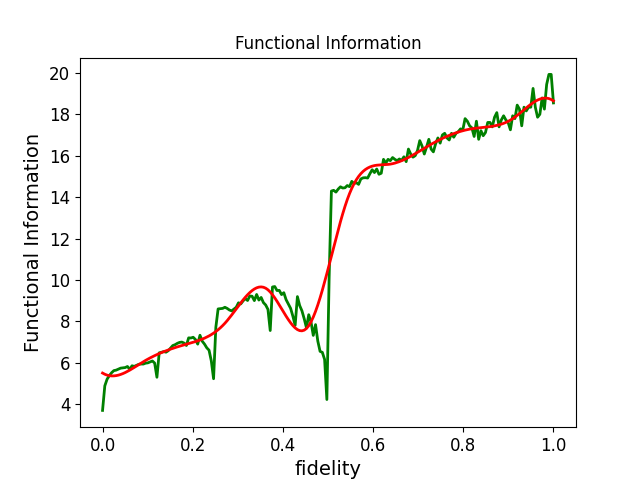}
        \caption{$n=4$ qubits}
    \end{subfigure}

    \begin{subfigure}{0.48\textwidth}
        \includegraphics[width=\linewidth]{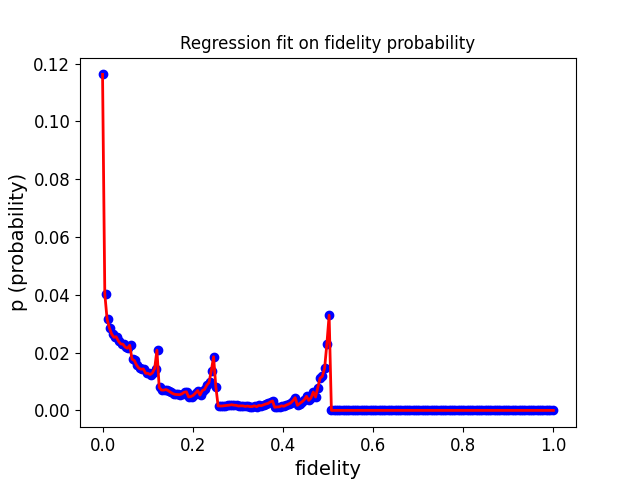}
        \caption{$n=5$ qubits}
    \end{subfigure}\hfill
    \begin{subfigure}{0.48\textwidth}
        \includegraphics[width=\linewidth]{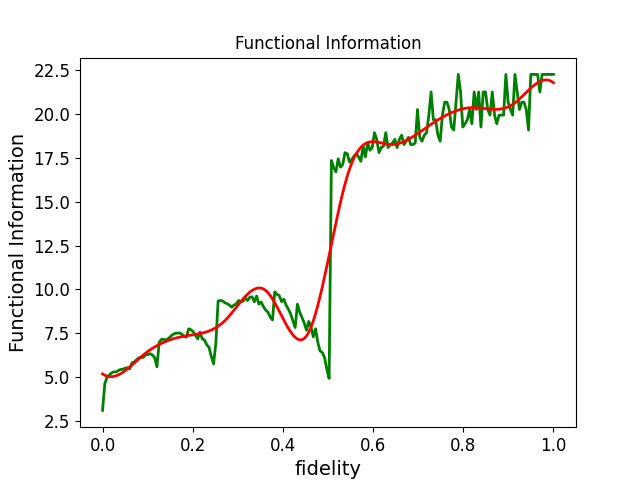}
        \caption{$n=5$ qubits}
    \end{subfigure}
 
    \caption{Final composition of functional information using regression of the fidelity probability and the application of $-log_2$. The red line is the smoothed function.}
    \label{fig:regression}
\end{figure}

Figure~\ref{fig:regression} present the regression results and the corresponding functional information curves. 
The red line in the plots indicates a secondary regression applied to smooth the functional information values, thereby providing a clearer view of the overall trend. To make this smooth regression, we use a spline-based regression model with Ridge regularization.  Thus, we can use the smoothed version of the functional information function.

As can be seen, the distributions and functional information follow the same pattern for 4 and 5 qubits. The same can be observed for 2 and 3 qubits, but we chose to show the results for 4 and 5 qubits. The distribution and the final QFI metric show that as the number of qubits increases, the functional information increases, which is expected given the different rarity distributions. The most noteworthy result is that circuits with fidelity equal to 1 tend to have less functional information than circuits that tend toward fidelity equal to 1 (e.g., 0.99, 0.98). This means that states that tend toward 1 are rarer than states with fidelity equal to 1. The same trend can be observed for 2, 3, 4, and 5 qubits, which were obtained empirically.
This empirical analysis demonstrates how quantum functional information can be directly estimated from large ensembles of random circuits and used to characterize the rarity of functional states such as entangled or GHZ-like states. Regarding the number of circuits and depth, through the correlation between fidelity and these variables, it is possible to observe that there is a negative Pearson correlation of approximately -0.639371 with the number of gates and -0.611654 with depth. This indicates that circuits with high fidelity tend to have a lower number of gates.

The probability is often extremely small in quantum systems due to the exponential size of the configuration space (e.g., \( 2^n \) states for \( n \) qubits or a vast number of possible gate sequences). A small probability like \( 10^{-6} \) is hard to compare intuitively across systems or tasks. The negative logarithm transforms this probability into a more interpretable quantity, measured in bits, representing the amount of information required to specify a functional configuration. For example, if \( P = 10^{-6} \), then \( I \approx 19.93 \) bits (\( -\log_2(10^{-6}) \approx -\log_2(2^{-19.93}) \)). This tells you the specificity of the functional state in a way that scales logarithmically, making it easier to compare systems with vastly different configuration spaces. In addition, the logarithmic scale amplifies differences in small probabilities. A change from \( F = 10^{-6} \) (\( I_f \approx 19.93 \) bits) to \( 10^{-7} \) (\( I \approx 23.25 \) bits) yields a clear 3.32-bit increase, highlighting the increased specificity required.  Finally, \( I \) bit-based units align with quantum information measures like entropy, facilitating integration with other metrics (e.g., entanglement entropy).

\section{Evolution of Quantum Functional Information}

This section explores the definition of Quantum Functional Information (QFI) through its application in a quantum circuit evolutionary process. The search is conducted by means of an Evolutionary algorithm (EA) \cite{Bhandari2024QuantumEvoCircuits, Liang2024SurveySurrogateAssisted}, which employs reproduction and mutation processes to maximize circuit functionality. Functionality is assessed according to the ability of a circuit to generate target states, with a particular focus on Bell and GHZ states, given their central role in entanglement and quantum information processing.  

The main objective of this experiment is to verify the effect of quantum functional information on the evolution of quantum circuits. Three evaluation scenarios are considered, each adopting a different objective function to guide the search:  

\begin{enumerate}
    \item \textbf{Fidelity as the objective function} – The EA is driven to maximize the fidelity between the evolved circuit’s output and the reference target state.
    \item \textbf{Quantum functional information as the objective function} – Instead of fidelity directly, the EA is guided by the functional information metric, which emphasizes the rarity of high-functionality states within the overall distribution of possible circuits.
\end{enumerate}

\subsection{Definition of the Evolutionary Algorithm}

The EA used in this study follows the classical structure of population-based metaheuristics. An initial population of randomly generated quantum circuits is produced. Each circuit is encoded as a sequence of quantum gates drawn from a predefined universal gate set. At each generation, the population undergoes variation through mutation (gate replacement, insertion, or deletion) and reproduction (selection and recombination of promising circuits). Selection is based on the defined objective functions, with elitism ensuring the preservation of the best candidates. The algorithm iterates until a stopping criterion, such as a maximum number of generations or convergence, is met.

The process can be summarized in the following steps:

\begin{itemize}
    \item \textbf{Initialization:} A population of random quantum circuits is generated using a mix of 1-qubit and 2-qubit gates.
    \item \textbf{Evaluation:} For each circuit, several metrics are computed:
    \begin{itemize}
        \item \emph{Fidelity (Fid)} with respect to the target state (Bell for $n=2$, GHZ otherwise).
        \item \emph{Average single-qubit entropy (Sv)}.
        \item \emph{Robustness (Rob)} defined as the ratio of noisy to ideal fidelity.
        \item \emph{Circuit depth}.
        \item \emph{Number of gates}.
    \end{itemize}
    
    \item \textbf{Selection:} Circuits are ranked by score, and the top 40\% are kept as elites.
    \item \textbf{Variation (Mutation):} New circuits are generated by mutating elites. Mutations include:
    \begin{itemize}
        \item Small perturbations of rotation angles.
        \item Random insertion of gates.
        \item Random deletion of gates.
    \end{itemize}
    \item \textbf{Replacement:} A new population is formed by elites and their mutated offspring.
    \item \textbf{Iteration:} Steps of evaluation, selection, and mutation are repeated for a fixed number of generations.
    \item \textbf{Output:} The algorithm returns the best circuit found, its metrics, the full evolutionary history, and all sampled individuals.
\end{itemize}

\noindent
The following pseudocode summarizes the evolutionary process:

\begin{tcolorbox}[colback=gray!5,colframe=black,title=Evolutionary Algorithm Pseudocode]
\begin{verbatim}
    function EVOLVE_CIRCUITS(n_qubits, pop_size, gens):
    # Initialization
    population ← [random_circuit(n_qubits) for i in 1..pop_size]

    for generation in 1..gens:
        scored ← []

        # Evaluation
        for each circuit in population:
            metrics ← QFI_metrics(circuit)   # Fid, Sv, Rob, depth, gates
            score, score_raw ← QFI_score(metrics)
            record (score, metrics, circuit)

        # Sort by score (higher is better, then by depth/gates)
        scored.sort(by = evo_sort_key)

        # Selection
        elites ← top 40% circuits from scored

        # Reproduction (mutation only)
        new_population ← elites
        while |new_population| < pop_size:
            parent ← random_choice(elites)
            child ← mutate(parent)
            add child to new_population

        population ← new_population

    # Final evaluation
    evaluate all population again
    return best_circuit, best_metrics, history, all_samples
\end{verbatim}
\end{tcolorbox}

\subsection{Other Metrics}

\subsubsection{Entropy of a Quantum Circuit}

Let $C$ be a quantum circuit acting on $n$ qubits, preparing the state $\rho_C$. To capture entanglement and correlation, we compute the von Neumann entropy of each single-qubit reduction $\rho_i = \tr_{\overline{i}} \rho_C$, with
\begin{equation}
Sv(\rho_i) = -\tr\!\left(\rho_i \log_2 \rho_i\right).
\end{equation}
The average entropy over qubits is then
\begin{equation}
\overline{Sv}(\rho_C) = \frac{1}{n} \sum_{i=1}^n S(\rho_i),
\end{equation}
which ranges from 0 (product state) to 1 (each qubit maximally mixed).  

To measure task-specific accuracy, we compute the fidelity with a pure target $\tau$. For $n=2$ we use the Bell state $\ket{\Phi^+} = (\ket{00}+\ket{11})/\sqrt{2}$, while for $n \geq 3$ the target is the GHZ state $\ket{\mathrm{GHZ}_n} = (\ket{0}^{\otimes n}+\ket{1}^{\otimes n})/\sqrt{2}$. The fidelity is
\begin{equation}
F(\rho_C,\tau) = \bra{\tau}\rho_C\ket{\tau}.
\end{equation}

\subsubsection{Robustness of a Quantum Circuit}

We define a robustness metric to capture the fraction of ideal fidelity preserved under noise. While we are not aware of an identical metric in prior literature, it is closely related to work such as Escofet et al. (2025) on fidelity under depolarizing noise \cite{Escofet2025FidelityDepolarizing}, and to robustness in control theory (e.g. Khalid et al. 2022) \cite{Khalid2022RobustnessFidelityControl}. Let $C$ be a quantum circuit acting on $n$ qubits and let $\tau$ be a fixed target state
(e.g., a Bell state for $n=2$ or a GHZ state for $n\ge 3$). Denote by
$\mathcal{U}_C(\cdot)$ the ideal (noise–free) channel implemented by $C$, and by
$\mathcal{N}_C(\cdot)$ the noisy channel obtained by composing $C$ with a gate–dependent
depolarizing noise model. For a fixed pure target $\tau = \lvert \psi_\tau \rangle
\langle \psi_\tau \rvert$, define the ideal and noisy fidelities
\begin{equation}
F_{\mathrm{ideal}}(C) \;=\; F\!\big(\mathcal{U}_C(\rho_0),\, \tau\big), 
\qquad
F_{\mathrm{noisy}}(C) \;=\; F\!\big(\mathcal{N}_C(\rho_0),\, \tau\big),
\end{equation}
where $\rho_0$ is the chosen input state (e.g., $\lvert 0\rangle^{\otimes n}$) and
$F(\rho,\sigma)$ is the  fidelity. Given a small constant $\varepsilon>0$ (used only to avoid division by zero in degenerate cases),
the robustness of $C$ is defined as
\begin{equation}
\label{eq:robustness}
R(C)
\;=\;
\min\!\left\{\,1,\;
\frac{F_{\mathrm{noisy}}(C)}{\max\!\big(F_{\mathrm{ideal}}(C),\,\varepsilon\big)}\right\}.
\end{equation}
In practice we set $\varepsilon\approx 10^{-9}$. When $F_{\mathrm{ideal}}(C)\ll \varepsilon$,
we use $R(C)=\min\{1,\,F_{\mathrm{noisy}}(C)\}$.

\begin{itemize}
\item $R(C)\in[0,1]$ by construction; $R(C)=1$ if the noisy fidelity matches (or exceeds) the
ideal fidelity (saturated by the cap).
\item If $F_{\mathrm{ideal}}(C)$ is high, then $R(C)\approx F_{\mathrm{noisy}}(C)/F_{\mathrm{ideal}}(C)$
quantifies how much of the \emph{useful} performance survives noise.
\item If $F_{\mathrm{ideal}}(C)\approx 0$, the fallback $R(C)\approx F_{\mathrm{noisy}}(C)$ avoids
numerical blow-up and preserves the $[0,1]$ range.
\end{itemize}

\subsection{Evolving Circuits Fidelity and QFI}

In this section, we will explore the effect of the QFI on the evolutionary process of circuit configurations (or quantum states). In the first case, pure fidelity is considered as the objective function, and in the second, QFI is used as the objective function. The results demonstrate the significant differences and how QFI is able to identify distinct patterns throughout the evolutionary processes. Thus, we name the two set of experiments as fidelity optimization and QFI optimization. In order to reach the conclusions presented here, we used a series of statistical analyses. We use the 4 qubits configuration.

\begin{figure}[h]
    \centering
    \begin{subfigure}{0.48\textwidth}
        \includegraphics[width=\linewidth]{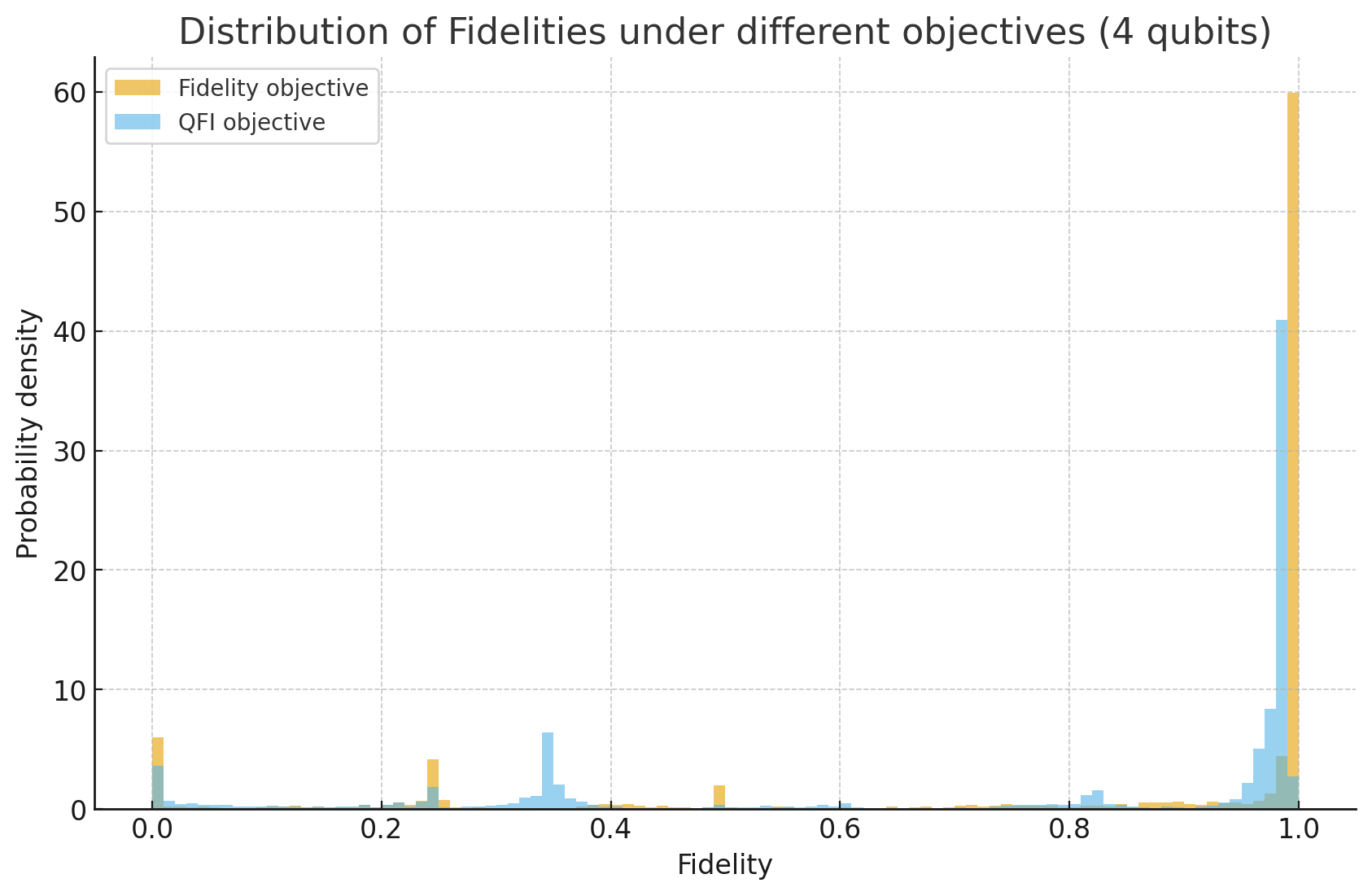}
        \caption{$n=2$ qubits}
    \end{subfigure}\hfill
    \begin{subfigure}{0.48\textwidth}
        \includegraphics[width=\linewidth]{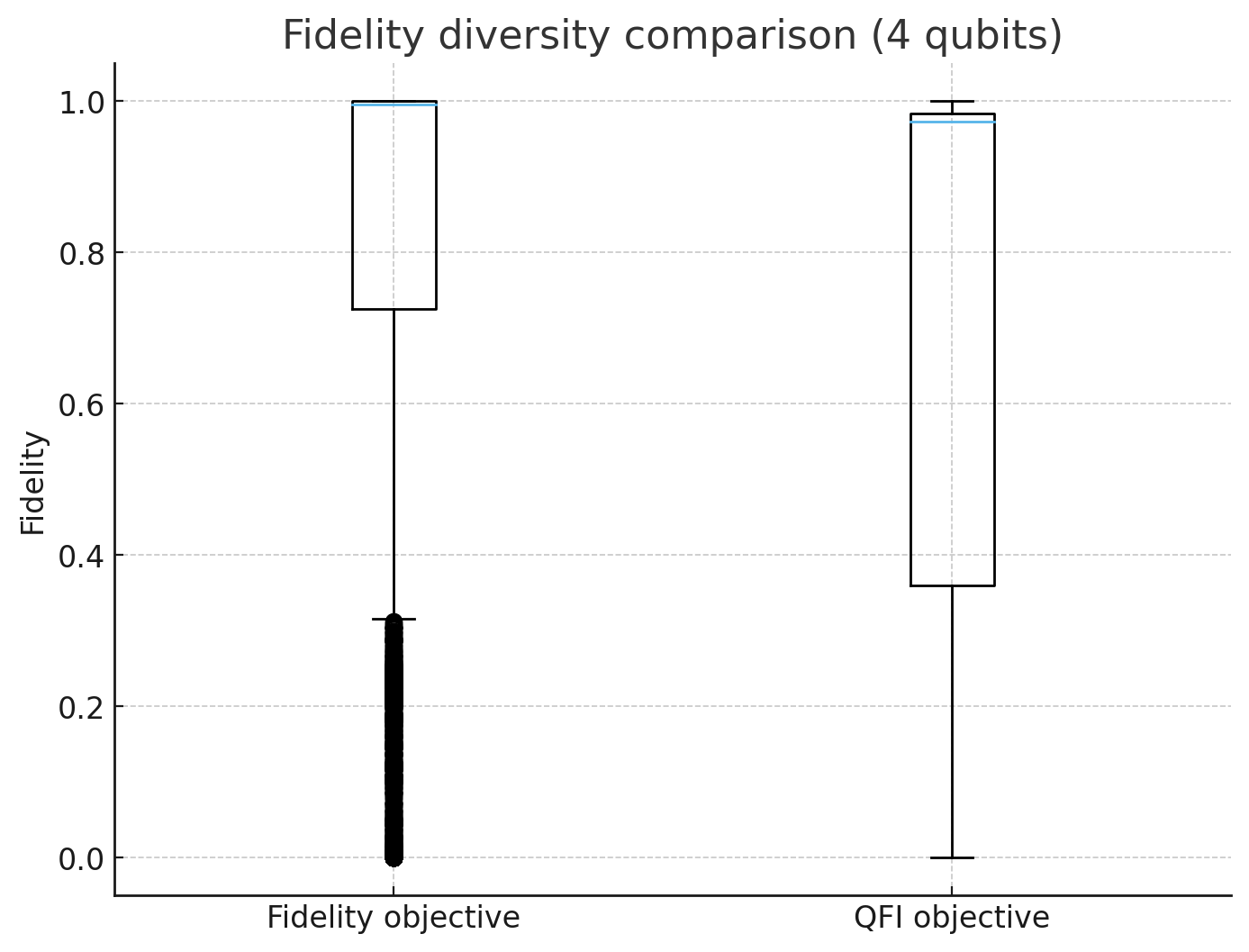}
        \caption{$n=3$ qubits}
    \end{subfigure}

    \caption{(a) Distribution of fidelity using QFI and fidelity as objective function. (b) Box plot comparison.}
    \label{fig:dist_boxplot_fid_qfi}
\end{figure}

Consider the Figure~\ref{fig:dist_boxplot_fid_qfi}. The analysis of fidelity distributions under the two optimization objectives reveals a marked contrast in behavior. When fidelity is used as the objective function, the evolutionary algorithm converges rapidly to solutions with fidelity values concentrated near $F \approx 1$. The resulting distribution is highly skewed, with a narrow interquartile range ($\mathrm{IQR} \approx 0.27$), reflecting premature convergence and limited diversity of solutions. 

In contrast, when quantum functional information (QFI) is employed as the optimization objective, the fidelity distribution is substantially broader. The interquartile range more than doubles ($\mathrm{IQR} \approx 0.64$), indicating that the algorithm maintains a wider variety of solutions across intermediate and high fidelity values. Although the mean fidelity is lower ($\overline{F} \approx 0.56$) compared to the fidelity objective run ($\overline{F} \approx 0.80$), the preservation of diversity is a key outcome: QFI optimization avoids collapse into a single high-fidelity basin and instead explores multiple functional regions of the search space. These results demonstrate that fidelity optimization promotes fast but narrow convergence, while QFI optimization enhances exploratory capacity, leading to a richer distribution of functionally significant circuits.

The boxplot comparison between the two optimization objectives highlights the impact of the chosen metric on convergence and diversity. For the fidelity objective, the distribution is extremely compressed near $F \approx 1$, with the median essentially coinciding with the maximum. This narrow box indicates that most solutions rapidly collapse into a single high-fidelity regime, leaving little variation within the population. In contrast, the QFI objective produces a markedly wider box, with the interquartile range more than twice that of the fidelity run. This demonstrates that QFI maintains a heterogeneous set of circuits, spanning from intermediate fidelity levels up to near-perfect states. Outliers are also more frequent, reinforcing that exploration of the solution space is broader.

\begin{figure}[h]
    \centering
    \includegraphics[width=0.7\linewidth]{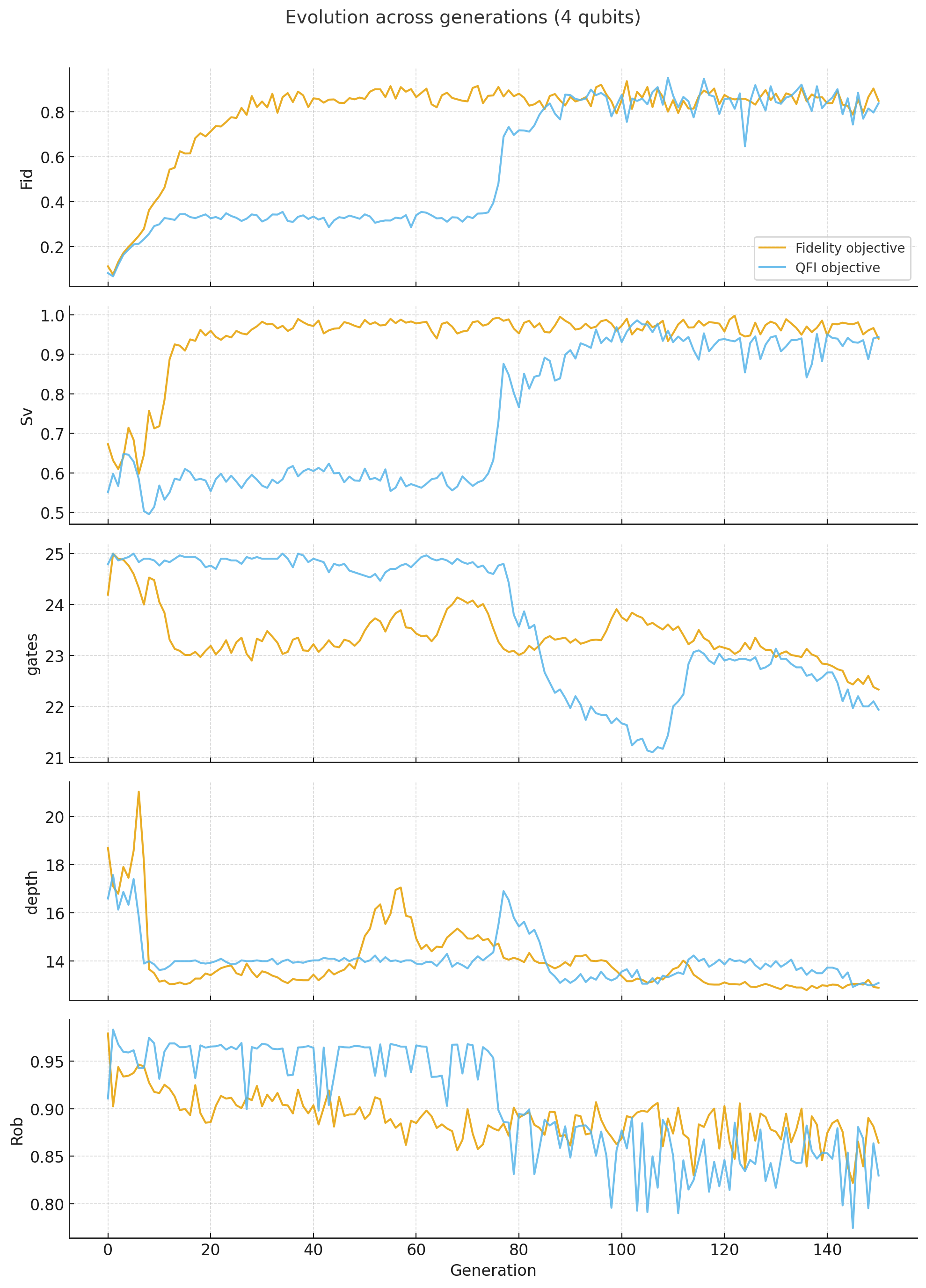}
    \caption{Distribution of fidelity using the QFI and the fidelity as objective function.}
    \label{fig:fid_qfi_gen_evolution}
\end{figure}

Now consider the Figure~\ref{fig:fid_qfi_gen_evolution}. The generational analysis reveals distinct optimization dynamics depending on the objective function. When fidelity is optimized directly, the algorithm exhibits a steep and rapid increase in performance, saturating at $F \approx 1$ within only a few generations. Once this plateau is reached, further evolution provides little improvement, and the population collapses into a narrow set of near-identical high-fidelity solutions.

In contrast, optimization guided by QFI displays a more gradual trajectory. Fidelity values rise steadily over successive generations but do not saturate as abruptly. Importantly, other metrics such as robustness, entropy, and circuit size evolve differently: robustness increases earlier and more consistently under QFI, producing solutions that are less sensitive to perturbations. The number of gates also tends to increase systematically, with QFI-optimized circuits frequently using the maximum gate  available, whereas fidelity-based optimization stabilizes at a slightly lower gate count. This reflects the fact that QFI encourages the exploration of more complex circuit structures in order to capture rare but functionally relevant states.

Entropy decreases under both objectives, but QFI tends to reduce entropy more strongly, yielding more structured quantum states. Circuit depth follows a similar pattern, with QFI maintaining deeper solutions on average. 

Overall, the evolutionary curves confirm that fidelity optimization drives fast exploitation but limited exploration, whereas QFI optimization balances exploration and exploitation, producing a broader set of functionally relevant circuits that are both more robust and structurally richer.

\begin{figure}[h]
    \centering
    \includegraphics[width=0.6\linewidth]{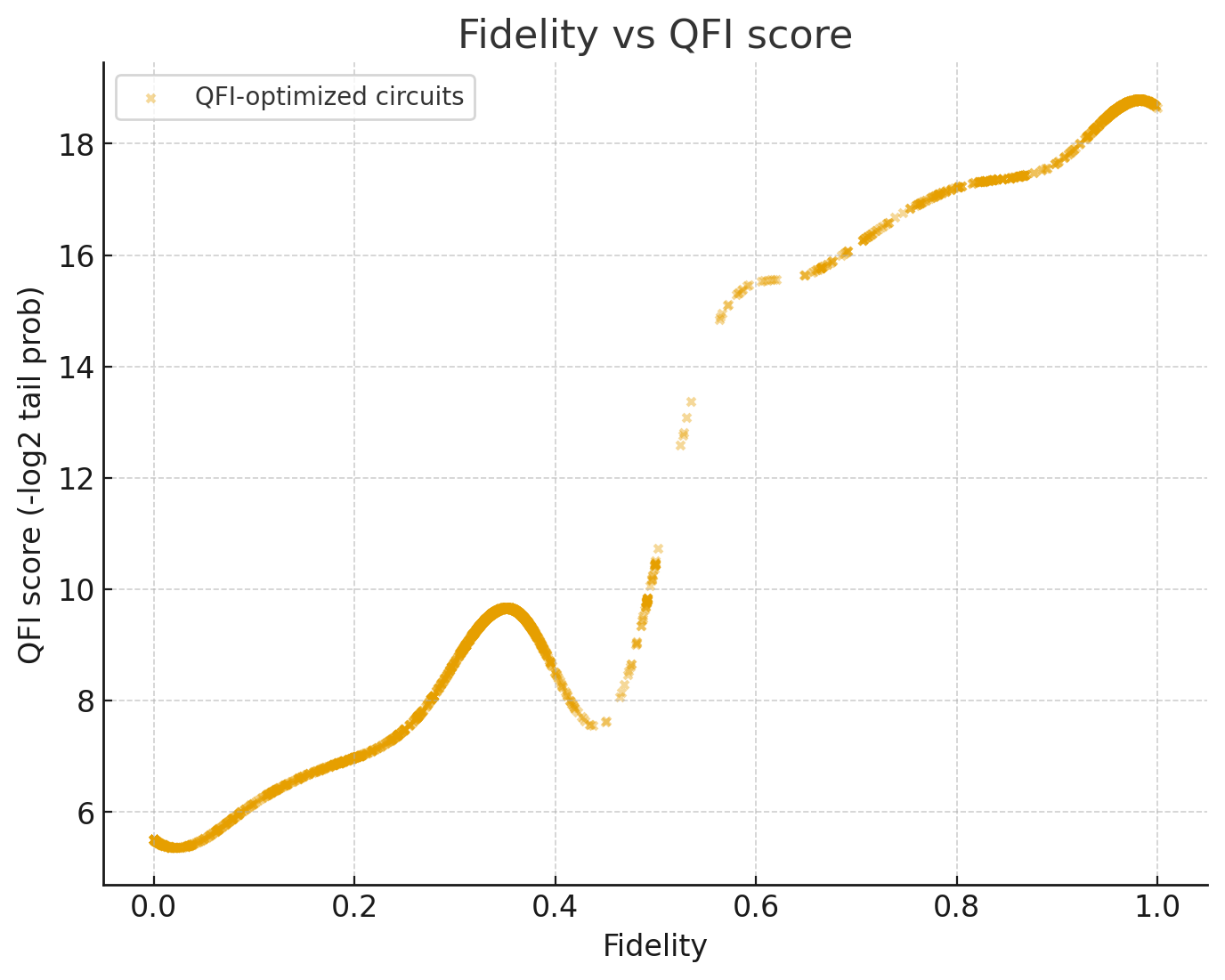}
    \caption{Analysis of fidelity vs QFI score as objective function.}
    \label{fig:fid_qfi_sampled}
\end{figure}

Considering the Figure~\ref{fig:fid_qfi_sampled}, we can see that maximum QFI scores cluster around high-but-not-perfect fidelities (0.95–0.98).The fidelity = 1 region (far right) doesn’t correspond to the highest QFI values, because once many circuits reach fidelity 1, those states are no longer rare. Instead, the QFI objective pulls the algorithm toward the region where circuits are both functional (good fidelity) and rare in the search space.

\begin{figure}[h]
    \centering
    \includegraphics[width=0.6\linewidth]{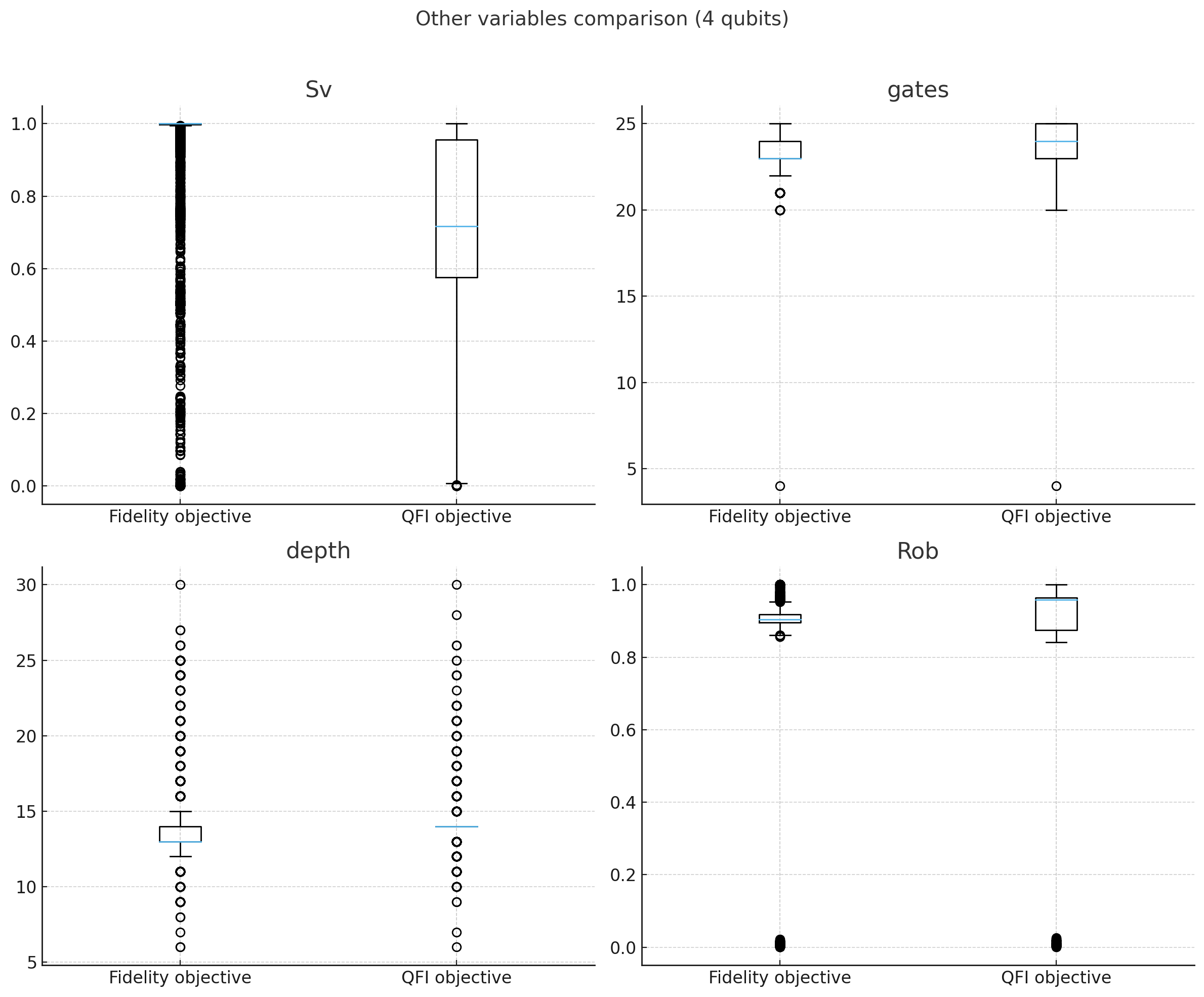}
    \caption{Distribution of other metrics using the QFI and the fidelity as objective function.}
    \label{fig:fid_qfi_boxplot_variables}
\end{figure}

Finally, we can see a view of four variables in boxplots in Figure~\ref{fig:fid_qfi_boxplot_variables}. The boxplots of entropy ($S_v$), number of gates, robustness, and depth reveal complementary differences between the two optimization strategies. 
For entropy, circuits evolved under fidelity optimization maintain higher values ($\overline{S_v} \approx 0.95$), whereas QFI-optimized circuits show a broader distribution with lower median entropy ($\overline{S_v} \approx 0.74$). This suggests that QFI favors more structured and less random states.

Regarding the number of gates, both strategies use circuits near the maximum allowed, but QFI optimization tends to push gate counts slightly higher, frequently saturating the maximum number. Fidelity optimization stabilizes at a lower and tighter distribution, indicating simpler circuits. A similar trend is observed for depth: fidelity-based runs stabilize around 14 layers with higher variability, while QFI maintains slightly deeper circuits with narrower dispersion.

Robustness exhibits the clearest contrast. Fidelity optimization yields a lower mean ($\overline{R} \approx 0.89$) with higher variance, whereas QFI optimization increases the median robustness ($\overline{R} \approx 0.90$) and reduces variability. This demonstrates that QFI tends to produce circuits that are not only structurally richer but also more stable under perturbations. These results confirm that fidelity optimization leads to simpler but less robust circuits, while QFI optimization sustains more complex architectures that balance depth and gate count with greater robustness and lower entropy.

\section{Conclusion}

The empirical analysis of random circuit ensembles confirmed that functional quantum states, such as Bell and GHZ configurations, are rare events within the Hilbert space. By sampling millions of circuits and estimating fidelity distributions, we demonstrated how Quantum Functional Information (QFI) can be directly computed as a rarity-based measure. The results revealed that states approaching—but not reaching—perfect fidelity tend to carry higher functional information than those at fidelity one, highlighting QFI’s ability to capture non-trivial patterns of rarity and functionality in quantum systems. 
The comparative analysis between fidelity-based and quantum functional information (QFI)-based optimization reveals clear and impactful differences in the behavior of the evolutionary algorithm. 

When fidelity is used as the sole objective, the search process converges rapidly to nearly perfect states ($F \approx 1$), but this convergence comes at the cost of diversity. The resulting circuits are simpler (fewer gates and shallower depth), but also less robust, with higher entropy values that reflect residual randomness in their structure. This outcome illustrates the risk of premature convergence: the algorithm exploits a narrow solution basin without maintaining exploratory capacity.

In contrast, optimization guided by QFI transforms the search dynamics. Instead of collapsing into a single high-fidelity, the population maintains a broad distribution of solutions across intermediate and high fidelities. This diversity is reflected in a significantly wider interquartile range of fidelities and in sustained exploration over generations. QFI-optimized circuits also display systematically higher robustness, lower entropy (indicating more structured states), and a tendency to employ more complex architectures by saturating the available number of gates. Importantly, the highest QFI values occur not at perfect fidelity, but slightly below it, confirming that QFI captures the combined attributes of \emph{functionality} and \emph{rarity} rather than accuracy alone.

The impact of these findings is twofold. First, QFI emerges as a  objective function that balances exploration and exploitation, avoiding the premature convergence inherent in fidelity-only optimization. Second, by promoting robustness, diversity, and structural richness, QFI-based evolution provides a pathway to discovering rare yet meaningful quantum circuits that may otherwise remain hidden in the vast search space. This demonstrates the potential of functional information as a guiding principle for evolutionary design in quantum computing. For example, in variational quantum algorithms \( I_f \) can serve as a cost function to optimize parameters, favoring configurations that maximize the fraction of functional states while minimizing specificity (i.e., making the algorithm more robust) \cite{Qi2024VQAreview}.

\section*{Acknowledgments}
We acknowledge the CISIA/PUCPR AI Center, and PPGCIS for computational resources. We also thank colleagues for discussions that shaped the concept of functional quantum information.

\bibliographystyle{unsrt}
\bibliography{REFS.bib}

\end{document}